\documentclass[vecphys]{svmult}

\usepackage{makeidx}         % allows index generation
\usepackage{graphicx}        % standard LaTeX graphics tool
\usepackage{multicol}        % used for the two-column index
\usepackage[bottom]{footmisc}% places footnotes at page bottom
\makeindex             % used for the subject index
\begin{document}

\title*{Resolving the Outer Disks and Halos of Nearby Galaxies}
%\titlerunning{Outer Disks and Halos of Nearby Galaxies}
% Use \titlerunning{Short Title} for an abbreviated version of
% your contribution title if the original one is too long
\author{Anil Seth\inst{1}\and
Roelof de Jong\inst{2}\and
David Radburn-Smith\inst{2}\and
%Eric Bell\and
%Tom Brown\and
%James Bullock\and
%Stephane Courteau\and
%Julianne Dalcanton\and
Harry Ferguson\inst{2}}
%Paul Goudfrooij\and
%Sherie Holfeltz\and
%Chris Purcell\and
%Dan Zucker}
%\authorrunning{Anil Seth, Roelof de Jong, and the GHOSTS team}
% Use \authorrunning{Short Title} for an abbreviated version of
% your contribution title if the original one is too long
\institute{Harvard-Smithsonian Center for Astrophysics,
\texttt{aseth@cfa.harvard.edu}
\and Space Telescope Science Institute}
%\and MPIA \and UC Irvine \and Queens U. \and U. Washington \and Cambridge}
%
% Use the package "url.sty" to avoid
% problems with special characters
% used in your e-mail or web address
%
\maketitle

\begin{figure}[t]
\centering
\includegraphics[width=10cm]{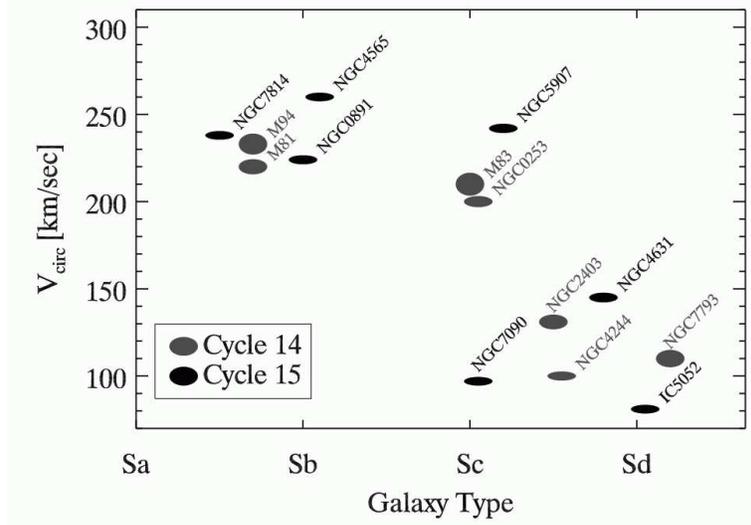}
\caption{The galaxy type vs. the circular velocity for the GHOSTS sample of galaxies.  Shape of the circle indicates the inclination of the galaxy.  Gray points were part of the Cycle 14 snapshot project, black points are part of the Cycle 15 large project.}
\label{Sethfig1}       % Give a unique label
\end{figure}

In a hierarchical merging scenario, the outer parts of a galaxy are a
fossil record of the galaxy's early history (e.g. \cite{bullock05}).
Observations of the outer disks and halos of galaxies thus provide a
tool to study individual galaxy histories and test formation theories.
Locally, an impressive effort has been made to understand the halo of
the Milky Way, Andromeda, and M33
(e.g. \cite{morrison00,gilbert06,ibata07} and contributions in this
volume).  However, due to the stochastic nature of halo formation, a
better understanding of this process requires a large sample of
galaxies with known halo properties.  The GHOSTS\footnote{GHOSTS $=$
Galaxy Halos, Outer disks, Streams, Thick disks, and Star clusters}
project (PI: R. de Jong) aims to characterize the halos and outer
portions of 14 nearby (D=4-14~Mpc) spiral galaxies using the Hubble
Space Telescope.  Figure~\ref{Sethfig1} shows the type, rotation
velocity and inclination of all 14 galaxies.  Detection of individual
stars in the outer parts of these galaxies enables us to study both
the morphological properties of the galaxies, and determine the stars'
metallicity and age.

\section{Disk and Halo Profiles}
\label{prosec}

\begin{figure}[t]
\centering
\includegraphics[height=5cm]{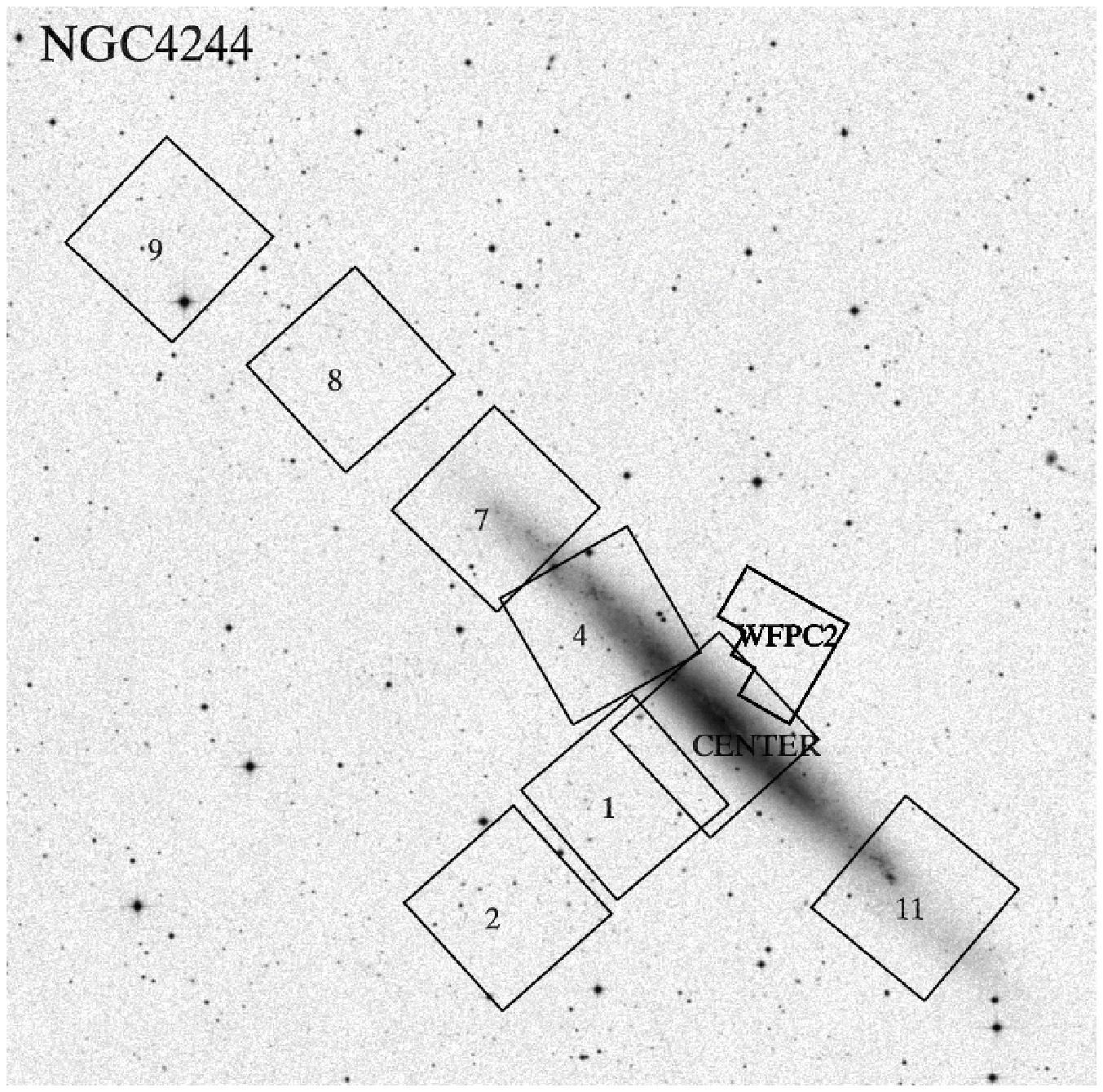}
\includegraphics[height=5.4cm]{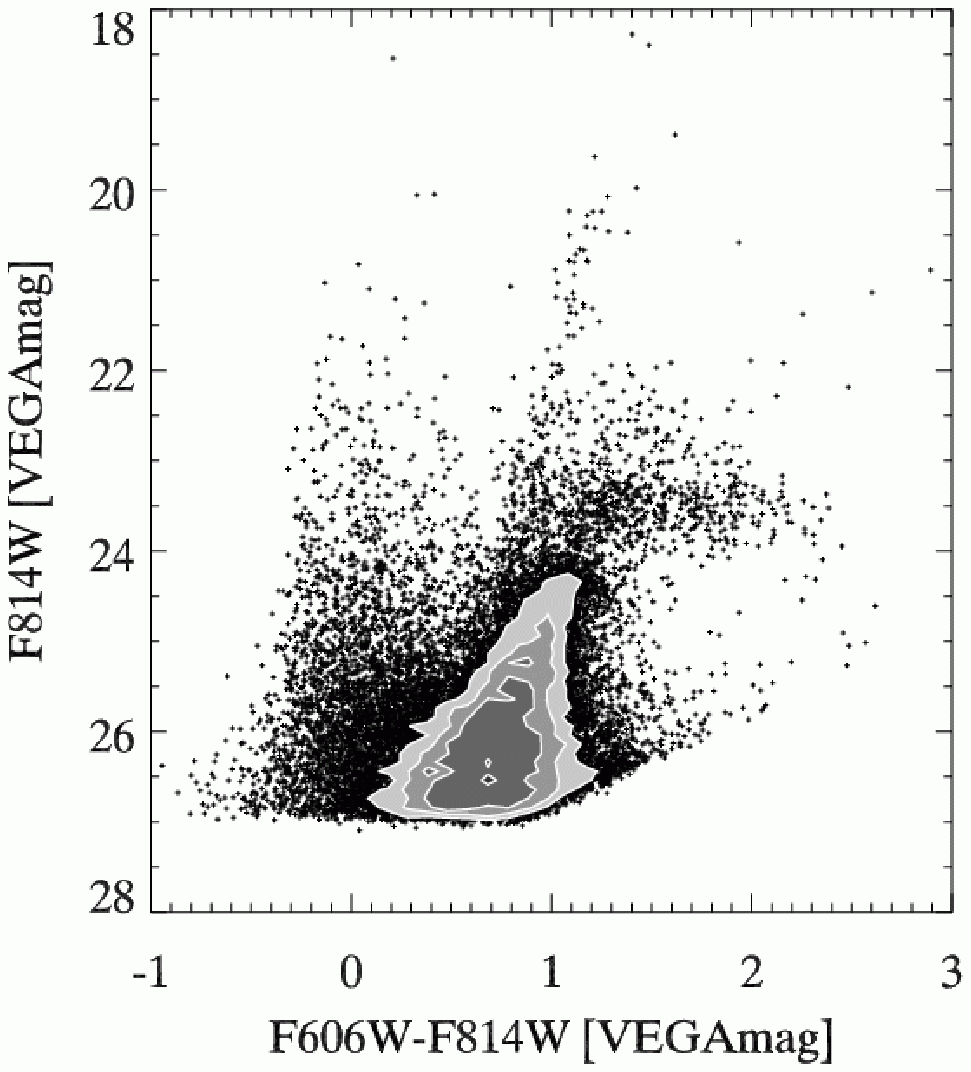}
\caption{{\it Left --} Numbered fields show the GHOSTS data for NGC~4244.  Other archival HST data is also shown.  {\it Right --} The color-magnitude diagram of field number 7 containing more than 20,000 stars.  The contours outline the RGB component, with the younger MS and AGB components creating the other features.}
\label{Sethfig2}       % Give a unique label
\end{figure}

The GHOSTS data consists of $\sim$6 ACS or WFPC2 fields in each
galaxy, primarily distributed along the major and minor axes (see
Fig.~\ref{Sethfig2}).  This areal coverage allows us to characterize
the shape of the outer disk and halo components, especially in the
eight edge-on galaxies in our sample where disk and halo components
can be easily separated. Using individual stars, we can trace out the
number density profile down to equivalent $i_{\rm AB}$-band surface
brightnesses of $\sim$31 mag/arcsec$^2$.  We discuss here two initial
results on the disk and halo profile of NGC~4244.

The edge-on galaxy NGC~4244 is a low mass ($V_{rot} \sim 100$km/sec),
SAcd galaxy, similar to local group spiral M33.  The GHOSTS data is
shown in Figure~\ref{Sethfig2}, with a CMD of one of the outer disk
fields showing the richness of the stellar populations detected in
our data.  In addition to the prominent red giant branch (RGB),
smaller numbers of young main sequence and helium burning stars (MS)
and asymptotic giant branch stars (AGB) are seen.  These populations
cover the entire history of the galaxy (see \cite{seth05b}).

Along the major axis, we find that the disk exhibits a strong
truncation at a radius of $\sim$420'' (see \cite{dejong07b} for
details).  Such truncations or breaks in the surface density profile
are commonly observed in disk galaxies, but we are able to resolve the
stellar populations across a truncation for the first time.
Interestingly, the break occurs in the same location for all of the
stellar populations from young to old.  Also, the break occurs
at the same radius for populations located above the midplane of the
galaxy as well.  These results show that the break radius in this
galaxy has been roughly constant over time, thus favoring dynamical
mechanisms for producing the break (e.g. \cite{debattista06}) versus
star formation threshold mechanisms (e.g. \cite{kennicutt89}).

The minor axis of NGC~4244 is also very interesting.  As noted in
\cite{seth05b}, the scale height of the old RGB population is higher
than the younger stellar populations.  With the GHOSTS data, we are
able to trace this RGB profile out to $\sim$10~kpc ($\sim$30 scale
heights) above the plane (see \cite{seth07a} for details).  The
exponential profile seen in our original data gives way to a slower
decline above $\sim$2.5~kpc.  If fit to an exponential, this more
diffuse component has a scale height similar to the scale length of
the main NGC~4244 disk, suggesting that we are detecting a spheroidal
halo.  Despite being very tenuous, this halo appears to
be more massive than the halo expected for such a low-mass galaxy
\cite{purcell07}.

\section{Metallicity and Age information}
\label{metsec}

The metallicity distribution of stars in the halos of galaxies
provides information on the mass of the galaxies which were
shredded to make the halo.  The color of individual RGB stars depends
primarily on their metallicity, and therefore is commonly used to
derive the metallicity distribution.

We have considered two different methods for constructing metallicity
distribution functions (MDFs). First, we interpolate each individual
star in our CMD onto a grid of isochrones assuming an age of 10~Gyr.
From this, we have found that a majority of the GHOSTS fields have
MDFs that peak at [Z]$\sim$-0.7.  This occurs in both fields which are
apparently dominated by stars in the outer disk and others which
appear to be halo-dominated.  

While the interpolation method is roughly correct, the photometric
errors translate into asymmetric errors in the metallicity and the
contribution of AGB stars below the tip of the RGB is ignored.  These
uncertainties particularly affect the metal-poor end, preventing us
from determining if these stars are present.

\begin{figure}[t]
\centering
\includegraphics[height=6cm]{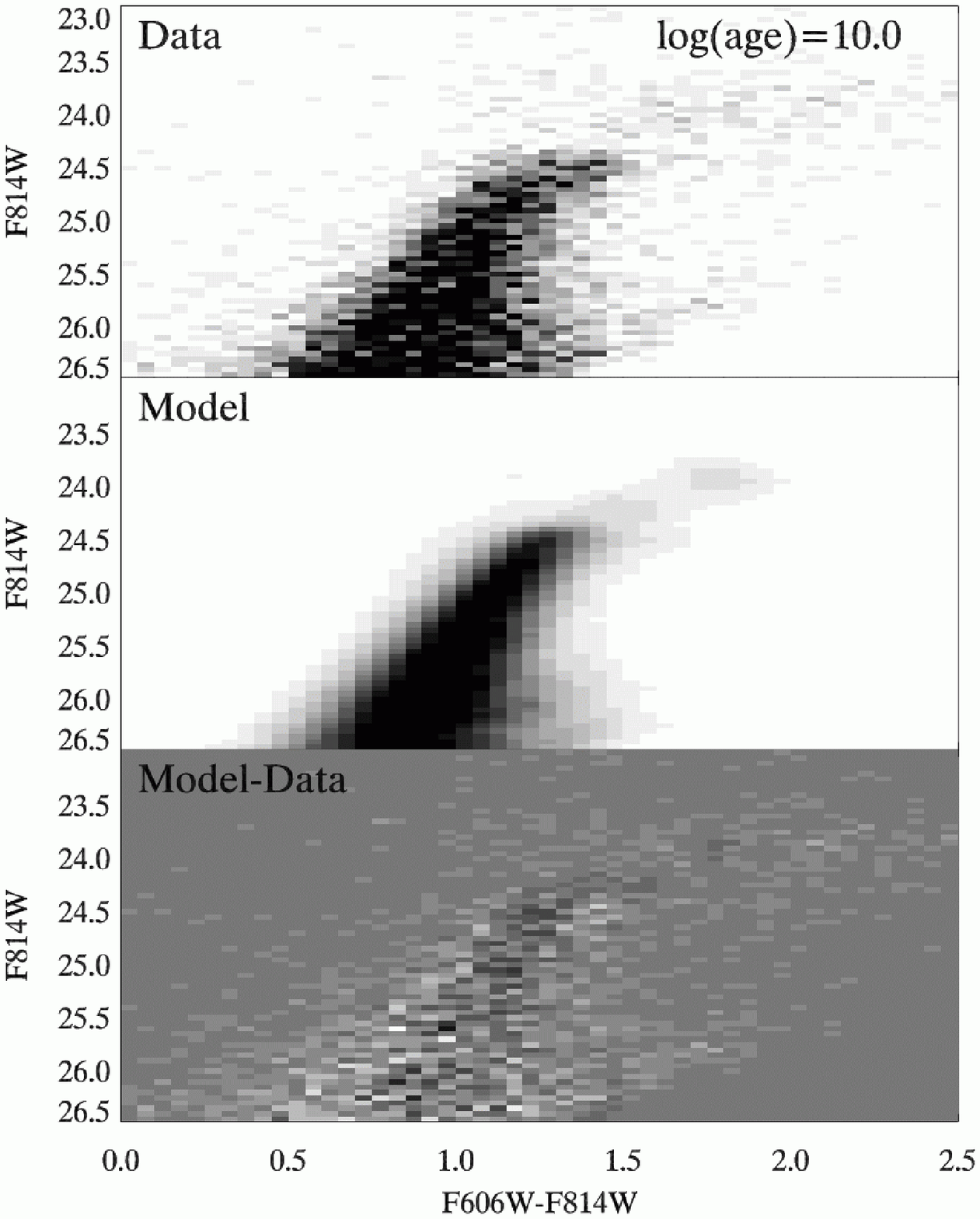}
\includegraphics[height=6cm]{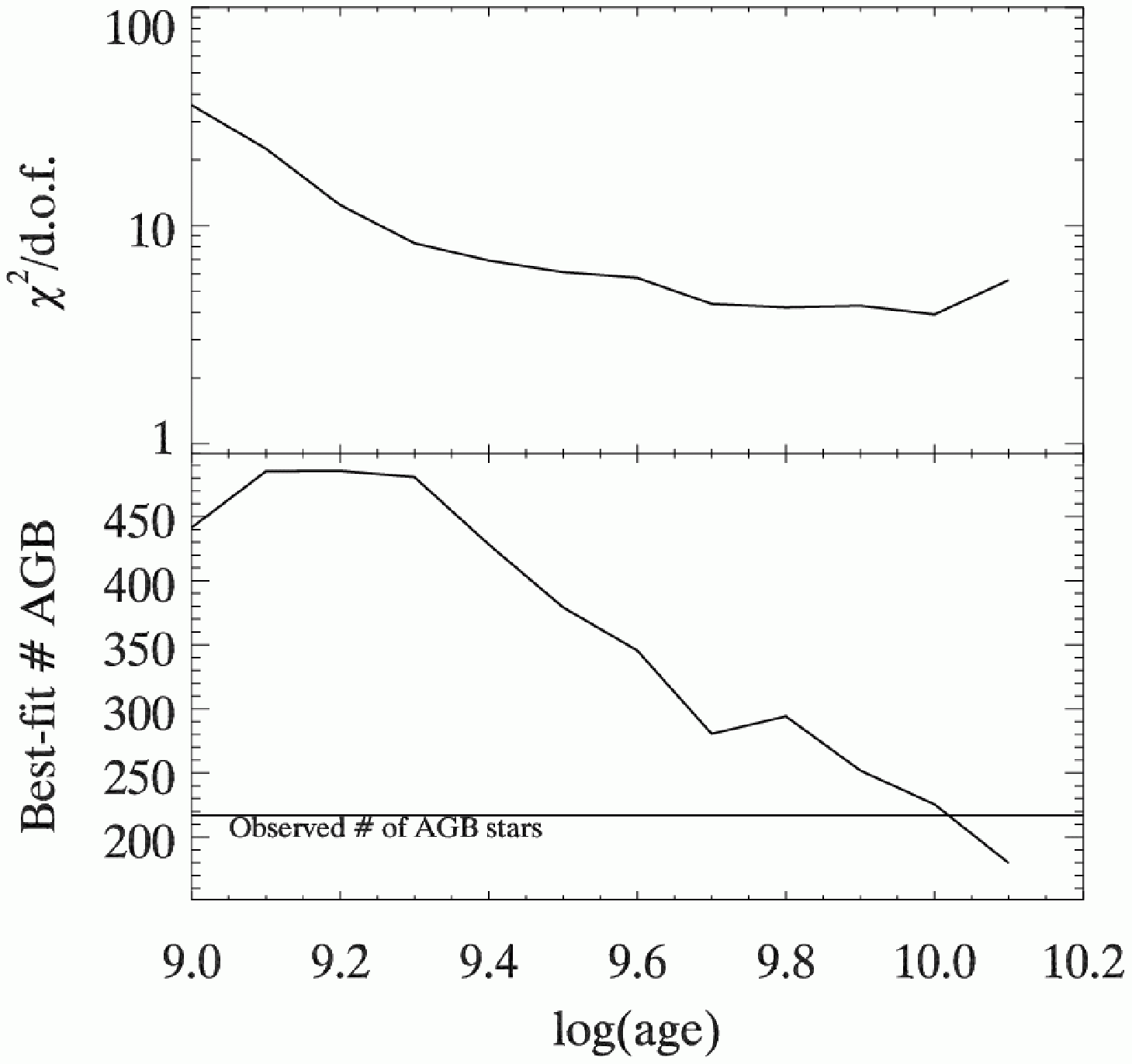}
\caption{{\it Left --} Hess diagrams of the M83 stream data, the best-fitting 10~Gyr model, and residuals.  {\it Right --} The best-fitting models' reduced $\chi^2$ and number of AGB stars versus age.  The AGB stars are significantly overproduced at ages younger than 5~Gyr.}
\label{Sethfig3}       % Give a unique label
\end{figure}

To try improving on this method we are using CMD fitting techniques in
which model CMDs of a given age (including both RGB and AGB stars) are
convolved with realistic errors.  As an example of this method, we
present results from our observations of a prominent stellar stream in
M83 at a projected distance of $\sim$25~kpc from the galaxy center.
Using the Starfish code \cite{harrisj02} and Padova models with updated AGB tracks \cite{marigo07},  we fit the CMD of this
field to a series of models at different ages.  The CMD of the stream
and the best-fitting model is shown in Figure~\ref{Sethfig3}.  This
model has a peak metallicity of [Z]=-0.5, suggesting the stream is
quite metal-rich.  Furthermore, the AGB stars provide us with the
possibility of constraining the age of the stream.  The right side of
Figure~3 shows the reduced $\chi^2$ of the best-fitting model vs. age.
Clearly models with ages larger than 5 Gyr are favored, primarily
because models with younger stars significantly overproduce AGB stars.

\subparagraph{Conclusions:}
The GHOSTS survey is providing information on the shape and
metallicity of the halos and outer disks of 14 nearby galaxies. We
have presented initial results on the major- and minor-axis profiles
of NGC~4244 and the metallicity and age of the stream in M83.
Analysis of the full sample of galaxies will provide strong
constraints on models of galaxy formation.

%
%\index{paragraph}
% Use the \index{} command to code your index words
%
%
% BibTeX users please use
% \bibliographystyle{apj}
% \bibliography{/Users/seth/allreferences.bib}
%
% Non-BibTeX users please follow the syntax
% the syntax of "referenc.tex" for your own citations
%\input{referenc}

%%%%%%%%%%%%%%%%%%%%%%%%%%%%%%%%%%%%%%%%%%%%%%%%%%%%%%%%%%%%%%%%%%%%%%  }

%%%%%%%%%%%%%%%%%%%%%%%%%%%%%%%%%%%%%%%%%%%%%%%%%%%%%%%%%%%%%%%%%%%%%%

\printindex
\end{document}